\documentclass[pra,aps,twocolumn, superscriptaddress,showpacs,tightenlines,amsmath]{revtex4}
\usepackage{epsfig}
\usepackage{amsmath,amssymb}
\usepackage{psfig}

\begin{document}

\title{Quantum Walks with Entangled Coins}
\author{S. E. Venegas-Andraca}
\affiliation{Centre for Quantum Computation, Clarendon Laboratory,
University of Oxford, Parks Road, Oxford OX1 3PU, United Kingdom}

\author{J. L. Ball}
\affiliation{Centre for Quantum Computation, Clarendon Laboratory,
University of Oxford, Parks Road, Oxford OX1 3PU, United Kingdom}

\author{K. Burnett}
\affiliation{Centre for Quantum Computation, Clarendon Laboratory,
University of Oxford, Parks Road, Oxford OX1 3PU, United Kingdom}

\author{S. Bose}
\affiliation{Department of Physics and Astronomy, University College
of London. Gower Street, London WC1E 6BT, United Kingdom.}

\begin{abstract}

We present a mathematical formalism for the description of
unrestricted quantum walks with entangled coins and one walker.
The numerical behaviour of such walks is examined when using
a Bell state as the initial coin state, two different coin operators, two
different shift operators, and one walker. We compare and contrast
the performance of these quantum walks with that of a classical
random walk consisting of one walker and two maximally correlated coins
as well as quantum walks with coins sharing different degrees of entanglement.

We illustrate that the behaviour of our walk with entangled coins can be very
different in comparison to the usual quantum walk with a single
coin. We also demonstrate that simply by changing the shift operator,
we can generate widely different distributions. We also compare the behaviour
of quantum walks with maximally entangled coins with that of quantum walks
with non-entangled coins. Finally,  we show that the use of different
shift operators on $2$ and $3$ qubit coins leads to different position probability
distributions in 1 and 2 dimensional graphs.


\end{abstract}

\maketitle

\section{introduction}

In recent years interest in the field of Quantum Walks has grown
hugely, motivated by the importance of Classical Random Walks in
Computer Science, as well as the advantages that Quantum Walks may
provide us with when compared to their classical counterparts.

Classical random walks are a fundamental tool in Computer Science
due to their use in the development of stochastic algorithms
\cite{motwani95}. In both theoretical and applied Computer Science,
stochastic algorithms may outperform any deterministic algorithm
built to solve certain problems. A notable example is that of the
best algorithm known so far for the solution of 3-SAT, a fundamental
problem in Computer Science which relies on random walks techniques
\cite{hofmeister02}.


So random walks are important elements of Computer Science.
Additionally, the recent development of Quantum Computation and
Quantum Information has revealed that the exploitation of inherently
quantum mechanical systems for computational purposes leads to a
number of significant advantages over purely classical systems. Thus
it is reasonable to expect that the study of random walks using
quantum mechanical systems may prove fruitful. The basic properties
of two kinds of quantum walks have already received a good deal of
attention: continuous (\cite{farhi98} - \cite{fedichkin04}) and
discrete (\cite{aharonov93} - \cite{knight04}) quantum walks. In
this paper we shall focus on discrete quantum walks.



Quantum walks are expected to play a major role in the field of
Quantum Algorithms. A number of benefits of employing such walks are
already known. In \cite{childs03}, Childs {\it et al} have built 
a quantum algorithm based on a continuous quantum walk to solve
the problem of traversing a graph $G$ with an exponential number of
vertices under a vertex connectivity arrangement. It was proved
in \cite{childs03} that traversing such a graph with their quantum algorithm
is exponentially faster than doing so with any classical algorithm.
Additionally, it has been proved by Ambainis {\it et al} that quantum
walks on a line spread quadratically faster than classical random walks
\cite{ambainis01}, while Kempe \cite{kempe03} proved that the hitting time
of a quantum walk on the hypercube is polynomial in the number of steps
of the walk (the classical counterpart takes exponential time). Also,
Shenvi {\it et al} \cite{shenvi02} showed that quantum walks can be used
to produce a quantum search algorithm. Finally, an algorithm based on
quantum walks to solve the problem of element distinctness
can be found in \cite{ambainis03}. An excellent summary of
the basics of quantum walks can be found in \cite{kempe04},
and a compendium of algorithmic applications of quantum walks
is presented in \cite{ambainis04}.

A discrete quantum walk is composed of two physical systems: a
walker and a coin (a detailed explanation of these two systems is
provided in Section III). The properties of quantum walks applying
multiple quantum coin operators (\cite{brunetal103} -
\cite{ambainiskempe04}) as well as decoherent coins (\cite{kendon02}
- \cite{brunetal203}) on a single walker have been extensively
studied.


However, the use of entanglement in quantum walks is less
well explored. A discussion on discrete quantum walks
using non-separable evolution operators and its effects on
the stardard deviation of resulting probability distributions
is given in \cite{mackay02}, followed by \cite{tregenna03}
where a more exhaustive study on non-separable operators
is provided. In \cite{du02}, Du {\it et al} proposed
an implementation of a continuous quantum walk on a circle,
and numerically showed that entanglement in the position states
shapes the position probability distribution. More recently,
a discussion concerning models of a quantum walk on a line
with two entangled particles as walkers is provided in 
\cite{paunkovic04}. A study of entanglement between coin
and walker in quantum walks on graphs is given in
\cite{carneiro05}, along with a generalization of
the quantum walk algorithm from \cite{childs03}. In \cite{endrejat05}
the authors analyse the relation between coin entanglement and
the mean position of the quantum walker for $3$ and $4$ qubit coins.
Finally, in \cite{abal05}, Abal {\it et al} have quantified the
entanglement between walker and coin generated by the shift operator
in a single coin-single walker quantum walk.

Our motivation to use entangled coins in quantum walks comes from
two sources. First, using entangled coins
$|c\rangle \in {\cal H}^n$ in quantum walks on graphs $G(V,E)$
with $\text{deg}(v_i) = m$ $\forall v_i \in V$ in which $n > m$, motivates
the employment of different shift operators and therefore
expands the dynamics of the quantum walk. In particular, in this
paper we use maximally entangled coins in quantum walks on an infinite
line along with shift operators with \lq \lq rest sites'', i.e.
states that allow the walker to stay at the current vertex. 
Indeed, it is also possible to introduce pairs of coins in
a classical random walk on an infinite line in order
to expand its dynamics, but that is at the expense of
varying the amount of correlation between the random
variables produced with the outcomes of corresponding
coins.


Second, an entangled coin comprised of two qubits, each residing in
$\mathcal{H}^2$, can be viewed as a single coin defined on
$\mathcal{H}^4$, and then appropriately partitioned. Indeed the
orthonormal basis $\{|00\rangle, |01\rangle, |10\rangle,
|11\rangle\}$ spans the space $\mathcal{H}^4$.  However, the
phenomenon of entanglement represents a supercorrelation between
possibly spacelike-separated subsystems of a total quantum system.
It is certainly feasible to generate entanglement between two
qubits and then separate them either for the purposes of an
experiment in the laboratory, for example to allow for individual
addressing of each qubit in some quantum information processing
experiment (such as implementing the Hadamard operator), or to
send them to opposite sides of the universe. This pre-existing
entanglement resource is created during a finite time period of
interaction and then the distinct subsystems can be separated to
arbitrary locations. However, a single four-level quantum system
cannot be physically broken and spatially separated into two
pieces so that each piece is subsequently subjected to local
operations. Partitioning a single coin into two entangled
subsystems is nevertheless equivalent to using two distinct
entangled coins, provided that the subsystems do not require to be
physically separated for practical purposes. Although the
mathematical description is the same, we choose to work with a
pair of entangled qubits. Generation of photonic entangled states,
for example by way of spontaneous parametric downconversion, or
entangled states in ion traps,  is already experimentally
achievable. As such, identifying the entangled coins as bipartite
states, rather than single an appropriately partitioned single
system residing in $\mathcal{H}^4$, is a natural choice to
highlight and motivate possible links to experiment.

Two specific physical implementations of quantum walks, namely
cavity-QED based \cite{sanders-bartlett} and ion-trap based
\cite{milburn-travaglione} motivate the scenario considered by us.
In both these implementations, two-level atoms serve as coins,
while a cavity mode \cite{sanders-bartlett} or a vibrational
mode \cite{milburn-travaglione} serve as the walker. Two atomic
qubits in an ion trap are already feasible, and have been prepared
in Bell states \cite{wineland} and there are several proposals for
entangling two atomic qubits in a cavity, such as \cite{beige-knight}.
These atoms can then be used as entangled coins with a common cavity mode
or the common ion trap vibrational mode acting as the walker controlled
by both these coins. While it is straightforward to treat the two atoms
as individual qubits during this process, it is rather difficult to do
entangling operations between them during the walk without using/affecting
the cavity or vibrational mode which is already acting as the walker.
Of course, a cavity mode or vibrational mode other than the one acting
as walker may be used to do entangling gates between the atoms, but this
is complicated. Moreover, in some cases, no method of accomplishing a direct
unitary entangling gate between two atoms may be present, and their initial
entanglement (needed for the walk considered here) may have been produced
using other mechanisms (such as decays and measurements \cite{beige-knight}).
Because of this inherent difficulty of doing an entangling gate between the
coins during a quantum walk, it is easier to imagine a scenario of
two entangled coin qubits rather than a single four dimensional coin.
Once two atoms have been trapped and entangled in a cavity, and this
has already been done for ion traps, the implementation of our scenario
is no more complex than a single coin quantum walk as the same global
fields can be applied to both atoms for the coin and the shift operations
(there is no need for addressing the atoms separately).

In this paper we shall discuss the behaviour of a quantum walk on an
infinite line (also called unrestricted quantum walk) with one coin
composed of two maximally entangled particles, and one walker. We
compare the performance of such a walk with that of a classical
random walk with one walker and two maximally correlated coins.
We also show that the use of different shift operators
on $2$ and $3$ qubit coins leads to different position probability
distributions in both one and two-dimensional graphs.

The idea behind correlated coins is simple. For a pair of correlated
coins $C_1$ and $C_2$ with corresponding outcomes ($H_1, T_1$) and
($H_2, T_2$) one expects that, after obtaining a certain outcome for
coin $C_1$, coin $C_2$ will produce its corresponding outcome {\it
according to a probability distribution defined by the degree of
correlation between both coins}.

For example, the behaviour of a maximally correlated pair of coins
would be the following: outcomes for coin $C_1$ would be given
according to a certain probability distribution. Let us suppose that
coin $C_1$ is unbiased, thus outcomes $H_1$ and $T_1$ may each occur
with equal probability. Now let us suppose that we get $H_1$ ($T_1$)
as outcome. Since the coin pair is maximally correlated, then the
outcome for coin $C_2$ will certainly be $H_2$ ($T_2$).

If the degree of correlation were less than maximal between coins
$C_1$ and $C_2$, then obtaining outcome $H_1$ for $C_1$ would imply
that the probability of getting $H_2$ as outcome for coin $C_2$
would not be unity. In fact the probability would scale as a
monotonically increasing function of the degree of correlation
between the coins.

Using correlated coins in classical random walks is straightforward.
For a classical random walk with a maximally correlated pair
of coins it is natural to assign the walker one step to the
right whenever the pair ($H_1, H_2$) (say) is the resulting outcome, and
one step to the left for the outcome ($T_1, T_2$) (say). In this case,
outcomes ($H_1, T_2$) and ($T_1, H_2$) have probability zero.

Indeed, we could enrich our classical random walk by allowing
coin outcomes $O_3 = (H_1, T_2)$ and $O_4 = (T_1, H_2)$. 
For example, one could use outcome $O_3$ to permit the walker to
remain in its current position or, alternatively, all four outcomes
could be used to perform a random walk with $1$ and $2$ steps to
the right and left, respectively. In particular, the introduction
of outcomes that allow rest states is a feature used to remove the
parity property of classical random walks, which consists of finding
the walker only in even (odd) positions in an even (odd) time step.
However, the introduction of outcomes $O_3$ and $O_4$ implies 
that the coin pair would no longer be maximally correlated.

In the following section we formally introduce a classical random
walk with a maximally correlated pair of coins. In Section III we
present a brief background introduction to unrestricted quantum
walks on a line with a single coin, followed by our results on
unrestricted quantum walks on a line with a maximally entangled
coin.

Our results show that probability distributions of quantum walks
with maximally entangled coins have particular shapes that are
highly invariant to changes in coin operators. These results are
then compared with those obtained for classical random walks with
maximally correlated coins. 


\section{Classical Random Walk with 2 Maximally Correlated Coins}

A classical result from stochastic processes states that, for an
unrestricted classical random walk starting at position $z_0=0$, the
probability of finding the walker at position $k$ after $n$ steps,
when with probability $p$ the walker takes a step to the right and
with probability $q=1-p$ takes a step to the left (i.e. tossing the
coin with probability $p$ of obtaining outcome $T$ and probability 
$q$ of obtaining outcome $H$), is given by

\begin{equation}\label{prob_dist_crw}
P_{ok}^{(n)} =
\binom{n}{\frac{1}{2}(k+n)}p^{\frac{1}{2}(k+n)}q^{\frac{1}{2}(n-k)}
\end{equation}
for $\frac{1}{2}(k+n) \in \{0, 1, \ldots, n\}$ and $0$ otherwise.

Tossing a pair of coins produces two discrete random variables
$C_1$ and $C_2$, and the correlation $\rho$ between these
two random variables is given by (\cite{grimmett91})
\begin{equation}
\rho(C_1,C_2) = { \text{Cov}(C_1,C_2) \over \sqrt{\text{Var}(C_1) \text{Var}(C_2)}}
\label{correlation}
\end{equation}
where Cov($X$,$Y$) and Var($X$) are the covariance and the variance
of the corresponding random variables. The function $\rho$ is bounded
by $-1 \leq \rho \leq 1$. $\rho(C_1,C_2) =0$ means that random variables
$C_1$ and $C_2$ are totally uncorrelated (i.e. $C_1$ and $C_2$ are independent),
whereas $\rho(C_1,C_2) =1$ means that random variables $C_1$ and
$C_2$ are maximally correlated. The case $\rho(C_1,C_2) =-1$ corresponds to
perfect anticorrelation.

\begin{figure}[h]
\epsfig{width=3in,file=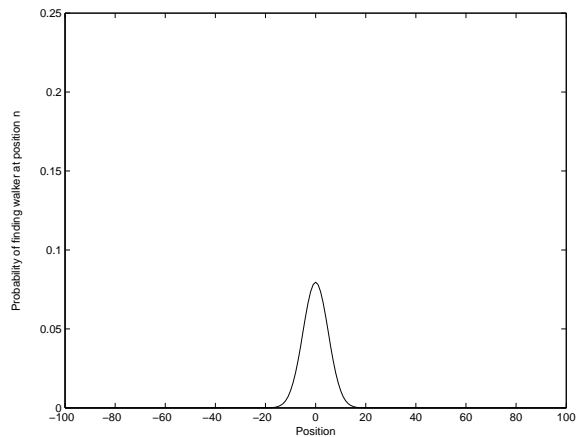} \caption{Plot of
$P_{ok}^{(n)} =
\binom{n}{\frac{1}{2}(k+n)}p^{\frac{1}{2}(k+n)}q^{\frac{1}{2}(n-k)}$
for $n=100$ and $p=\frac{1}{2}$. The probability of finding the
walker in position $k=0$ is equal to 0.0795. Only probabilities
corresponding to even positions are shown, as odd positions have
probability equal to zero.}\label{binomial_08}
\end{figure}

Now consider a classical random walk that has a maximally correlated
pair of coins, i.e. $\rho(C_1,C_2)=1$. Also suppose that the first
coin $C_1$ is unbiased. Then, as explained in the previous section,
the only two outcomes allowed for this coin pair are $O_1 = (H_1,
H_2)$ or $O_2 = (T_1, T_2)$. If $O_1$ allows the walker to move one
step to the left and $O_2$ allows the walker to move one step to the
right, it is then clear that using such a coin pair in a classical
random walk would produce a probability distribution equal to that
of Eq. (\ref{prob_dist_crw}), with $p = \frac{1}{2}$. A plot of Eq.
(\ref{prob_dist_crw}) with number of steps $n=100$ and
$p=\frac{1}{2}$ is provided in Fig. (\ref{binomial_08}) for the
purpose of comparison with results presented in Section III.

As can be seen in Fig.(\ref{binomial_08}), the use of maximally
correlated unbiased coins in classical random walks is not
different with respect to a classical random walk with a single
unbiased coin, as the probability distributions from both kinds
of classical random walks are exactly the same.

In the following sections we shall compare the results obtained
by the computation of classical random walks with maximally
correlated (classical) coins with those of quantum walks
with maximally entangled (quantum correlated) coins.

\section {Quantum Walks with Entangled Coins}

\subsection{Review of Quantum Walks on an Infinite Line}

We now review the mathematical structure of a quantum walk on a line
with one coin and a single walker, with a view to using this
structure to construct a model for unrestricted quantum walks on the
line with a maximally entangled coin.

The main components of a quantum walk on a line are a walker, a
coin, evolution operators for both walker and coin, and a set of
observables:

\noindent\emph{\textbf{Walker and Coin:}} The walker is a quantum
system living in a Hilbert space of infinite but countable dimension
${\cal H}_p$. It is customary to use vectors from the canonical
(computational) basis of ${\cal H}_p$ as \lq\lq position sites" for
the walker. So, we denote the walker as $|\text{position}\rangle \in
{\cal H}_p$ and affirm that the canonical basis states $|i
\rangle_p$ that span ${\cal H}_p$, as well as any superposition 
of the form $\sum_{i} \alpha_i|i\rangle_p$ subject to $\sum_i|\alpha_i|^2 = 1$,
are valid states for $|\text{position}\rangle$. The walker is usually
initialized at the \lq origin', i.e.
$|\text{position}\rangle_{\text{initial}} = |0\rangle_p$.

The coin is a quantum system living in a 2-dimensional Hilbert space
${\cal H}_c$. The coin may take the canonical basis states $|0\rangle$
and $|1\rangle$ as well as any superposition of these basis states.
Therefore $|$coin$\rangle$ $\in {\cal H}_c$ and a general normalized
state of the coin may be written as
$|$coin$\rangle$ $= a |0\rangle_c + b |1\rangle_c$, where
$|a|^2 + |b|^2 = 1$.

The total state of the quantum walk resides in ${\cal H}_t = {\cal
H}_p \otimes {\cal H}_c$. Only product states of ${\cal H}_t$ have
been used as initial states, that is, $|
\psi\rangle_{\text{initial}} =
|\text{position}\rangle_{\text{initial}} \otimes
|\text{coin}\rangle_{\text{initial}}$.

\noindent\emph{\textbf{Evolution Operators:}}  The evolution of a
quantum walk is divided into two parts that closely resemble the
behaviour of a classical random walk. In the classical case, chance
plays a key role in the evolution of the system. This is evident in
the following example: we first toss a coin (either biased or
unbiased) and then, depending on the coin outcome, the walker moves
one step either to the right or to the left.

In the quantum case, the equivalent of the previous process is to
apply an evolution operator to the coin state followed by a
conditional shift operator to the total system. The purpose of the
coin operator is to render the coin state in a superposition, and
the randomness is introduced by performing a measurement on the
system after both evolution operators have been applied to the total
quantum system several times.

Among coin operators, customarily denoted by $\hat{C}$, the Hadamard
operator
\begin{equation}
{\hat H} = {1 \over \sqrt{2}} (|0\rangle_{cc} \langle 0| + |0
\rangle_{cc} \langle 1| + |1\rangle_{cc} \langle 0| - |1\rangle_{cc}
\langle 1|)
\label{hadamard_single}
\end{equation}
has been extensively used.

For the conditional shift operator use is made of a unitary operator
that allows the walker to go one step forward if the accompanying
coin state is one of the two basis states (e.g. $|0\rangle$), or one
step backwards if the accompanying coin state is the other basis
state ($|1\rangle$). A suitable conditional shift operator has the
form
\begin{eqnarray}
{\hat S} = |0\rangle_{cc} \langle 0| \otimes \sum_i |i+1
\rangle_{pp} \langle i| \nonumber\\
+ |1\rangle_{cc} \langle 1| \otimes \sum_i |i-1 \rangle_{pp} \langle
i|.
\label{shift_single}
\end{eqnarray}

Consequently, the operator on the total Hilbert space is ${\hat U} =
{\hat S}.({\hat C} \otimes \mathbb{{\hat I}}_p)$ and a succint
mathematical representation of a quantum walk after $n$ steps is $|
\psi \rangle = ({\hat U})^n |\psi\rangle_\text{initial}$, where
$|\psi\rangle_{\text{initial}} = |\text{position}
\rangle_{\text{initial}} \otimes |\text{coin}
\rangle_{\text{initial}}$.

\noindent\emph{\textbf{Observables:}}  The advantages of quantum
walks over classical random walks are a consequence of interference
effects between coin and walker after several applications of ${\hat
U}$. However, we must perform a measurement at some point in order
to know the outcome of our walk. To do so, we define a set of
observables according to the basis states that have been used to
define coin and walker.

In order to extract information from the composite quantum system,
we first perform a measurement on the coin using the observable
\begin{equation}
{\hat M}_c = \alpha_0 |0\rangle_{cc}\langle 0| + \alpha_1
|1\rangle_{cc}\langle 1|.
\end{equation}
A measurement must then be performed on the position states of the
walker by using the operator
\begin{equation}
{\hat M}_p = \sum_i a_i |i\rangle_{pp}\langle i|.
\end{equation}

\subsection{Mathematical Structure of Quantum Walks on an Infinite Line Using
a Maximally Entangled Coin}

As before, the elements of an unrestricted quantum walk on a line
are a walker, a coin, evolution operators for both coin and walker,
and a set of observables. We shall provide a detailed description of
each element motivated by the previous subsection.

\noindent\emph{\textbf{Walker and Coin:}}  The walker is, as in the
unrestricted quantum walk with a single coin, a quantum system
$|\text{position}\rangle$ residing in a Hilbert space of infinite
but countable dimension ${\cal H}_P$. The canonical basis states $|i
\rangle_P$ that span ${\cal H}_P$, as well as any superposition 
of the form $\sum_{i} \alpha_i|i\rangle_p$ subject to
$\sum_i|\alpha_i|^2 = 1$, are valid states for the walker.
The walker is usually initialized at the \lq origin' i.e.
$|\text{position}\rangle_0 = |0\rangle_P$.

The coin is now an entangled system of two qubits i.e. a quantum
system living in a 4-dimensional Hilbert space ${\cal H}_{EC}$. 
We denote coin initial states as $|\text{coin}\rangle_0$. Also, 
we shall use the following Bell states as coin initial states 
\begin{subequations}
\begin{equation}\label{bell1}
|\Phi^+\rangle = \frac{1}{\sqrt{2}} (|00\rangle + |11\rangle)
\end{equation}
\begin{equation}\label{bell2}
|\Phi^-\rangle = \frac{1}{\sqrt{2}} (|00\rangle - |11\rangle)
\end{equation}
\begin{equation}\label{bell3}
|\Psi^+\rangle = \frac{1}{\sqrt{2}} (|01\rangle + |10\rangle)
\end{equation}
\end{subequations}
which are maximally entangled pure bipartite states with reduced von
Neumann entropy equal to unity. We shall examine the consequences of
employing such maximally entangled states by comparing the resulting
walks with those resulting from using maximally correlated coins in
classical random walks. The Bell singlet state $|\Psi^-\rangle =
\frac{1}{\sqrt{2}} (|01\rangle - |10\rangle)$ is not employed as an
entangled coin as it is left invariant when the same local unitary
operator is applied to both coins.

The total initial state of the quantum walk resides in the Hilbert space
${\cal H}_T = {\cal H}_P \otimes {\cal H}_{EC}$ and has the form
\begin{equation}
|\psi\rangle_0 = |\text{position}\rangle_0 \otimes
|\text{coin}\rangle_0
\end{equation}

\noindent\emph{\textbf{Entanglement measure:}}
In order to quantify the degree of entanglement of the coins
used in this paper, we shall employ the reduced von Neumann entropy
measure. For a pure quantum state $| \psi \rangle$ of a composite
system $AB$ with $\text{dim}(A) = d_A$ and $\text{dim}(B) = d_B$,
let $|\psi \rangle = \sum_{i=1}^d \alpha_i |i_A\rangle |i_B\rangle$,
($d$ = min($d_A$,$d_B$), $\alpha_i \geq 0$ and $\sum_{i=1}^d \alpha_i^2 = 1$) be its
Schmidt decomposition. Also, let $\rho_A = \text{tr}_B(|\psi\rangle \langle \psi|)$
and $\rho_B = \text{tr}_A(|\psi\rangle \langle \psi|)$ be the reduced density
operators of systems $A$ and $B$ respectively. {\it The entropy of entanglement}
$E(|\psi\rangle)$ is the von Neumann entropy of the reduced density operator \cite{bennett96}

\begin{equation}
E(|\psi\rangle) = S(\rho_A) = S(\rho_B) = - \sum_{i=1}^d \alpha_i^2 \log_2 (\alpha_i^2).
\label{von_neumann}
\end{equation}
$E$ is a monotonically-increasing function of the entanglement present in the system
$AB$. A non-entangled state has $E=0$.
States $|\psi\rangle \in {\cal H}^d$ for which $E(\psi) = d$ are called {\it maximally
entangled states} in $d$ dimensions. In particular, note that
for those quantum states described by Eqs. (\ref{bell1}),
(\ref{bell2}) and (\ref{bell3}) 
$E(|\Phi^+\rangle) = E(|\Phi^-\rangle) = E(|\Psi^+\rangle) = 1$,
i.e. these states are maximally entangled.

\noindent\emph{\textbf{Evolution Operators:}}  The evolution
operators used are more complex than those for quantum walks with
single coins. As in the single coin case, the only requirement
evolution operators must fulfil is that of unitarity.

Let us start by defining evolution operators for an entangled coin.
Since the coin is a bipartite system, its evolution operator is
defined as the tensor product of two single-qubit coin operators:
\begin{equation}
\hat{C}_{EC} = \hat{C} \otimes \hat{C}
\end{equation}
For example, we could define the operator $\hat{C}_{EC}^H$ as the
tensor product $\hat{H}^{\otimes 2}$:
\begin{equation}\label{hadamard_tensor}
\begin{split}
\hat{C}_{EC}^H = & {1 \over 2}(|00\rangle \langle 00| + |01\rangle
\langle 00| + |10\rangle
\langle 00| + |11\rangle \langle 00| \\
& + |00\rangle \langle 01| - |01\rangle \langle 01| + |10\rangle
\langle 01| -
|11\rangle \langle 01| \\
& + |00\rangle \langle 10| + |01\rangle \langle 10| - |10\rangle
\langle 10| -
|11\rangle \langle 10| \\
& + |00\rangle \langle 11| - |01\rangle \langle 11| - |10\rangle
\langle 11| + |11\rangle \langle 11|).
\end{split}
\end{equation}
An alternative bipartite coin operator is produced by computing the
tensor product $\hat{Y}^{\otimes 2}$ where $\hat{Y}
=\frac{1}{\sqrt{2}}(|0\rangle \langle 0| + i|0\rangle \langle 1|+
i|1\rangle \langle 0| + |1\rangle \langle 1|)$, namely
\begin{equation}\label{y_tensor}
\begin{split}
\hat{C}_{EC}^Y = & {1 \over 2} ( |00\rangle \langle 00| +
i|01\rangle \langle 00| + i|10\rangle
\langle 00| - |11\rangle \langle 00| \\
& + i|00\rangle \langle 01| + |01\rangle \langle 01| - |10\rangle
\langle 01|
+ i|11\rangle \langle 01| \\
& + i|00\rangle \langle 10| - |01\rangle \langle 10| + |10\rangle
\langle 10|
+ i|11\rangle \langle 10| \\
& - |00\rangle \langle 11| + i|01\rangle \langle 11| + i|10\rangle
\langle 11| + |11\rangle \langle 11|).
\end{split}
\end{equation}

Both coin operators are fully separable, thus any entanglement
in the coins is due to the initial states used.

The conditional shift operator $\hat{S}_{EC}$ necessarily allows the
walker to move either forwards or backwards along the line,
depending on the state of the coin. The operator

\begin{equation}\label{shift_operator_two_spins}
\begin{split}
{\hat S}_{EC} = |00\rangle_{cc} \langle 00| \otimes \sum_i |i+1
\rangle_{pp}
\langle i| \\
+ |01\rangle_{cc} \langle 01| \otimes \sum_i |i\rangle_{pp} \langle i| \\
+ |10\rangle_{cc} \langle 10| \otimes \sum_i |i\rangle_{pp} \langle i|\\
+ |11\rangle_{cc} \langle 11| \otimes \sum_i |i-1 \rangle_{pp}
\langle i|
\end{split}
\end{equation}
embodies the stochastic behaviour of a classical random walk with a
maximally correlated coin pair. It is only when both coins reside in
the $|00\rangle$ or $|11\rangle$ state that the walker moves either
forwards or backwards along the line; otherwise the walker does not
move.

Note that $\hat{S}_{EC}$ is one of a family of valid definable shift
operators. Indeed, it might be troublesome to identify a classical
counterpart for some of these operators: their existence is uniquely
quantum-mechanical in origin. One such alternative operator is
\begin{equation}\label{shift_operator_s2}
\begin{split}
{\hat S}_{EC}' = |00\rangle_{cc} \langle 00| \otimes \sum_i |i+2
\rangle_{pp}
\langle i| \\
+ |01\rangle_{cc} \langle 01| \otimes \sum_i |i+1\rangle_{pp} \langle i| \\
+ |10\rangle_{cc} \langle 10| \otimes \sum_i |i-1\rangle_{pp} \langle i|\\
+ |11\rangle_{cc} \langle 11| \otimes \sum_i |i-2 \rangle_{pp}
\langle i|.
\end{split}
\end{equation}

The total evolution operator has the structure ${\hat U}_T = {\hat
S_{EC}}.({\hat C_{EC}} \otimes \mathbb{{\hat I}}_p)$ and a succint
mathematical representation of a quantum walk after $N$ steps is 
$|\psi \rangle = ({\hat U}_T)^N |\psi\rangle_0$, where $|\psi\rangle_0$
denotes the initial state of the walker and the coin.

\noindent\emph{\textbf{Observables:}} The observables defined here
are used to extract information about the state of the quantum walk
$| \psi \rangle = ({\hat U}_T)^N |\psi\rangle_0$.

We first perform measurements on the coin using the observable
\begin{equation}
\begin{split}
{\hat M}_{EC} = \beta_{00} |00\rangle_{cc}\langle 00| + \beta_{01}
|01\rangle_
{cc}\langle 01|\\
\beta_{10} |10\rangle_{cc}\langle 10|+\beta_{11}
|11\rangle_{cc}\langle 11|.
\end{split}
\end{equation}
Measurements are then performed on the position states using the
operator
\begin{equation}
{\hat M}_P = \sum_j b_j |j\rangle_{PP}\langle j|.
\end{equation}

With the purpose of introducing the results presented
in the rest of this paper we compare in Table 1 the 
actual position probability values for a classical random walk
on an infinite line (Eq. (\ref{prob_dist_crw})), and a
quantum walk with initial state $|\Phi^+\rangle = {1 \over \sqrt{2}}(|00\rangle + |11\rangle)$
and coin and shift operators given by Eqs. (\ref{hadamard_tensor})
and (\ref{shift_operator_two_spins}), respectively.

{\bf Table 1.} 
Position Probability values for classical random walk and quantum walk
\\
\[\begin{array}{|c|c|c|c|c|c|c|c|}
\hline
\text{{\bf Classical}} & $-3$  & $-2$ & $-1$ & $ 0$ & $ 1$ & $ 2$ & $ 3$\\
\hline
\text{ Step } 0        & $0$   & $0$  & $0$  & $1$  & $0$ & $0$ & $0$\\
\hline
\text{ Step } 1        & $0$   & $0$  & $1/2$  & $0$  & $1/2$ & $0$ & $0$\\
\hline
\text{ Step } 2        & $0$  & $2/8$ & $0$ & $4/8$ & $0$ & $2/8$ & $0$\\
\hline
\text{ Step } 3        & $4/32$  & $0$ & $12/32$ & $0$ & $12/32$ & $0$ & $4/32$\\
\hline
\end{array}\]

\[\begin{array}{|c|c|c|c|c|c|c|c|}
\hline
\text{{\bf Quantum}}   & $-3$   & $-2$   & $-1$   & $ 0$   & $ 1$   & $ 2$   & $ 3$\\
\hline
\text{ Step } 0        & $0$    & $0$    & $0$    & $1$    & $0$    & $0$    & $0$\\
\hline
\text{ Step } 1        & $0$    & $0$    & $1/2$  & $0$    & $1/2$  & $0$    & $0$\\
\hline
\text{ Step } 2        & $0$    & $1/8$  & $2/8$  & $2/8$  & $2/8$  & $1/8$  & $0$\\
\hline
\text{ Step } 3        & $1/32$ & $6/32$ & $5/32$ & $8/32$ & $5/32$ & $6/32$ & $1/32$\\
\hline
\end{array}\]

\subsection{Results for Quantum Walks on an Infinite Line Using
a Maximally Entangled Coin}



\begin{figure}[h]
\epsfig{width=3in,file= 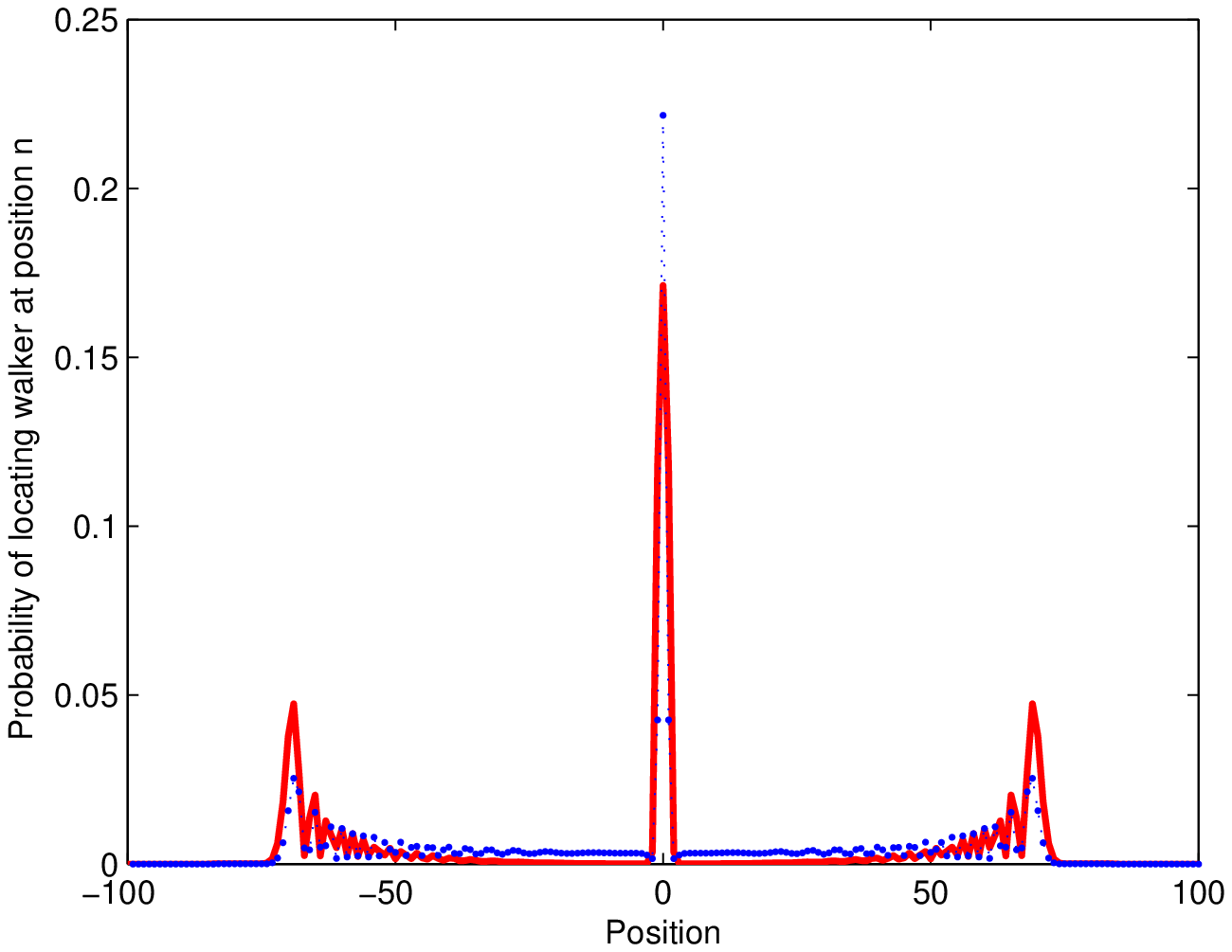}
\caption{For both plots, coin initial state is
$|\Phi^+\rangle = \frac{1}{\sqrt{2}} (|00\rangle + |11\rangle)$
and the number of steps is $100$. Coin operators for red 
and dotted blue plots are given by Eq. (\ref{hadamard_tensor})
and Eq. (\ref{y_tensor}) respectively.} \label{figure2}
\end{figure}

\begin{figure}[h]
\epsfig{width=3in,file= 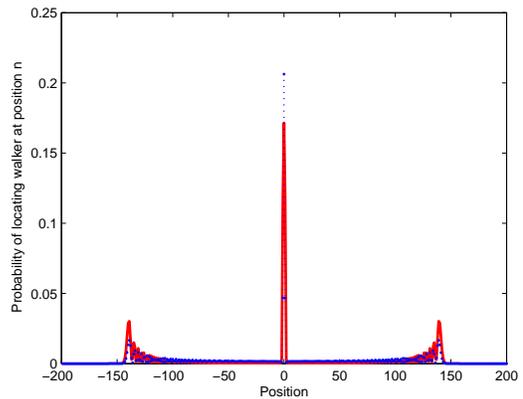}
\caption{For both plots, coin initial state is
$|\Phi^+\rangle = \frac{1}{\sqrt{2}} (|00\rangle + |11\rangle)$
and the number of steps is $200$. Coin operators for red
and dotted blue plots are given by Eq. (\ref{hadamard_tensor})
and Eq. (\ref{y_tensor}) respectively.} \label{figure3}
\end{figure}

In order to investigate the properties of unrestricted quantum walks
with entangled coins, we have computed several simulations using
bipartite maximally entangled coin states described by Eqs.~
(\ref{bell1}), (\ref{bell2}) and (\ref{bell3}), and coin evolution
operators described by Eqs. (\ref{hadamard_tensor}) and
(\ref{y_tensor}). In all cases, the initial position state of the
walker corresponds to the origin, i.e. $|\text{position}\rangle_0 =
|0\rangle$. The shift operator employed is, with the exception of
Fig.~({\ref{figure8}), that of Eq.~(\ref{shift_operator_two_spins}).

Let us first discuss the quantum walks whose graphs are shown in
Fig.~(\ref{figure2}). The initial entangled coin state is given by
Eq.~(\ref{bell1}) and the number of steps is $100$. For the red plot 
in Fig.~(\ref{figure2}) the coin operator is given by Eq.~(\ref{hadamard_tensor}),
while for the dotted blue plot in the same Fig.~(\ref{figure2})
the coin operator is that of Eq.~(\ref{y_tensor}).

The first notable property of these quantum walks is that, unlike
the classical case in which the most probable location of the walker
is at the origin and the probability distribution has a single peak,
in the quantum case a certain range of very likely positions about
the position $|0\rangle$ is evident but in addition there are a
further two regions at the extreme zones of the walk in which it is
likely to find the particle. This is the \lq three peak zones'
property of the shift operator defined in this way. 
The \lq three peak zones' property could mean an additional
advantage of quantum walks over classical random walks.

We also note that the probability of finding the walker in the most
likely position, $|0\rangle$, is much higher in the quantum case
($\sim 0.171242$ in red plot of Fig.~(\ref{figure2}) and $\sim 0.221622$
in dotted blue plot of Fig.~(\ref{figure2})) than in the classical
case $(\sim 0.0795)$. Incidentally, we find that the use of different
coin initial states maintains the basic structure of the probability
distribution, unlike the quantum walk with a single coin in which
the use of different coin initial states can lead to different probability
distributions (\cite{aharonov01}, \cite{bach04} and \cite{yamasaki03}) .


The position probability distributions shown in Fig.~(\ref{figure2})
could embody some advantages when used in an appropriate
application framework. For example, let us suppose we want
to design algorithms whose purpose is to find the wrong values
in a proposed solution of a problem (a concrete case is
to find all wrong binary values assigned to the initial
conditions of the algorithm used in \cite{schoning99} to
find a solution to the 3-SAT problem.) 
We use a $100$-steps classical random walk (Fig.~(\ref{binomial_08}))
to design algorithm $C$ and a $100$-steps quantum walk with maximally
entangled coins (red plot of Fig.~(\ref{figure2})) to design algorithm $Q$. 

Depending on the actual number of wrong values, the probability 
distribution of red plot of Fig. (\ref{figure2}) could help to
make algorithm $Q$ faster than algorithm $C$. For example, if
the number of wrong values is in the range $40$ - $70$,
the probability of finding the quantum walker of 
Fig. (\ref{figure2}) is much higher than finding the classical
walker of Fig. (\ref{binomial_08}). Actual probability values
(note the differences in orders of magnitude)  are shown in Table 2.
It must be noted that employing a quantum walk
on a line with a single coin for building algorithm $Q$ would also 
produce higher probability values than a classical random walk in
those positions shown in Table 2, thus the choice of quantum walk
could depend on some other factors like implementation feasibility.
\\
\\
{\bf Table 2.} Position Probabilities for
Classical and Quantum Walkers
\\
\begin{tabular}{|l|l|l|l|}
\hline
Position & Classical Walker          & Quantum Walker\\
\hline
40       & $2.31 \times 10^{-5}$    & $1.80 \times 10^{-3}$\\
\hline
50       & $1.91 \times 10^{-7}$    & $1.50 \times 10^{-3}$\\
\hline
60       & $4.22 \times 10^{-10}$     & $1.03 \times 10^{-2}$\\
\hline
70       & $1.99 \times 10^{-13}$    & $3.78 \times 10^{-2}$\\
\hline
\end{tabular}
\\
\\
\\

\begin{figure}[h]
\epsfig{width=3in,file=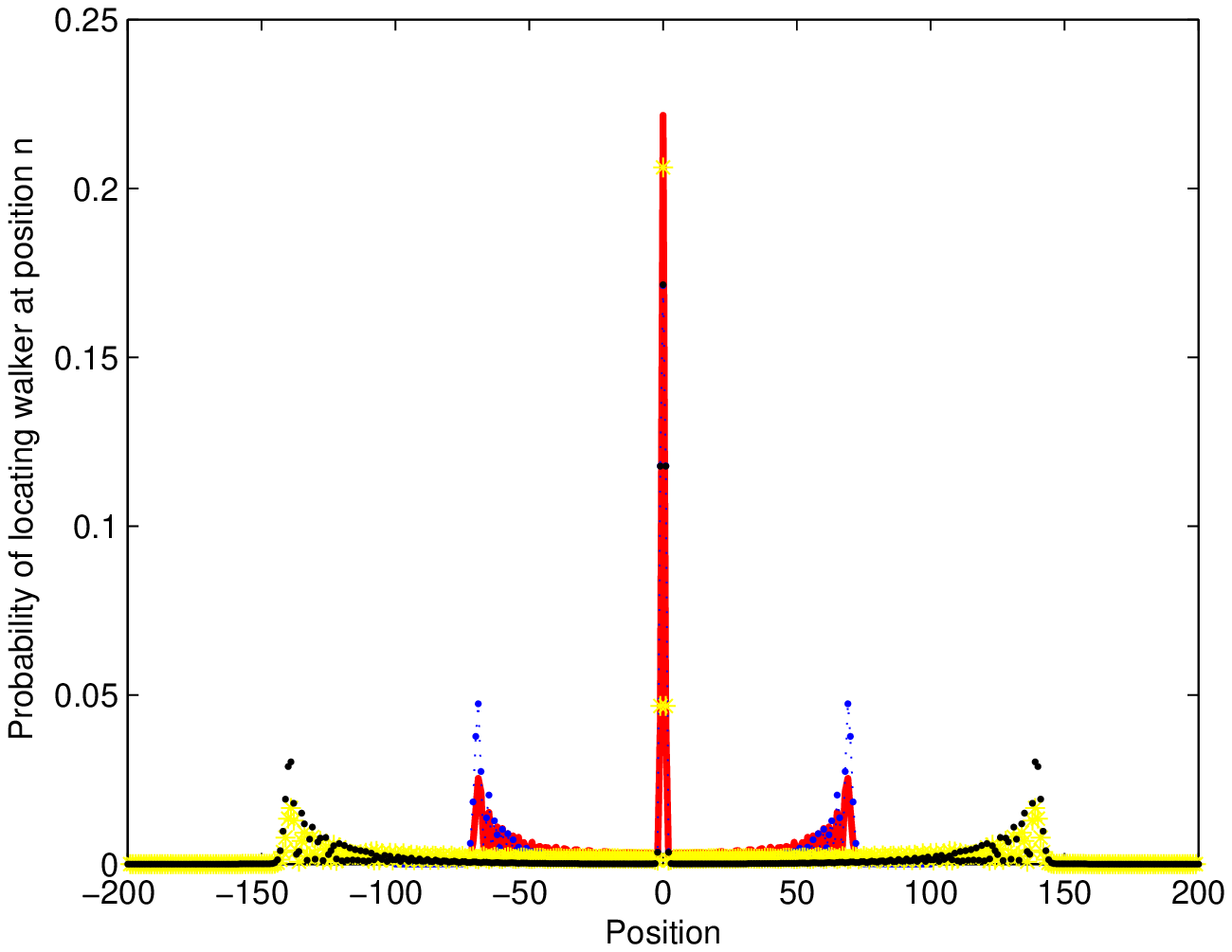}
\caption{Coin initial state is $|\Phi^-\rangle = \frac{1}{\sqrt{2}} (|00\rangle -
|11\rangle)$. Number of steps is $100$ for
red and dotted blue plots and $200$ for starred
yellow and black dotted plots. Coin operator for 
red and starred yellow plots is given by
Eq. (\ref{hadamard_tensor}), while coin operators
for dotted blue and dotted black plots is given
by Eq. (\ref{y_tensor}) respectively.} \label{figure4}
\end{figure}

\begin{figure}[h]
\epsfig{width=3in,file=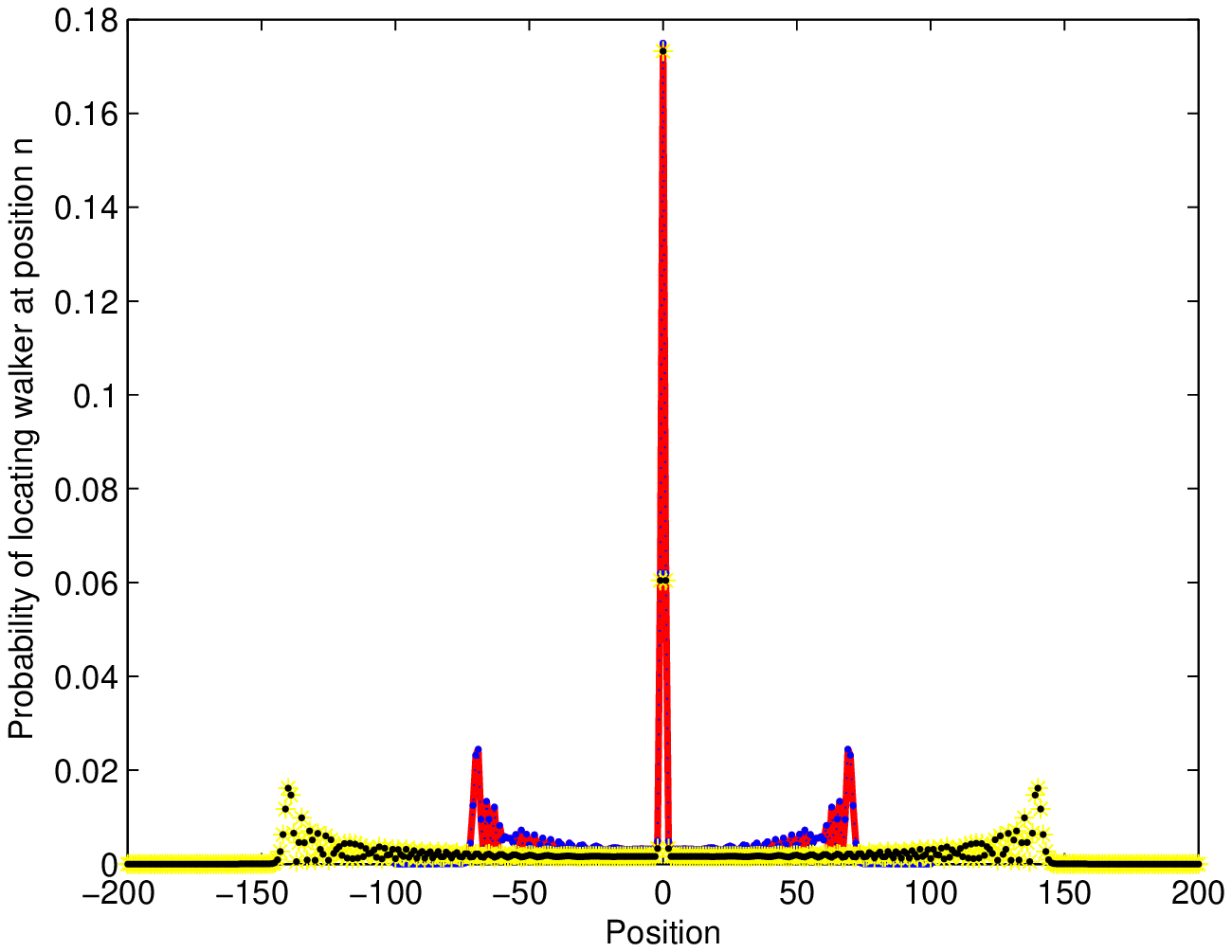}
\caption{Coin initial
state is $|\Psi^+\rangle = \frac{1}{\sqrt{2}} (|01\rangle +
|10\rangle). $Number of steps is $100$ for
red and dotted blue plots and $200$ for starred
yellow and black dotted plots. Coin operator for 
red and starred yellow plots is given by
Eq. (\ref{hadamard_tensor}), while coin operators
for dotted blue and dotted black plots is given
by Eq. (\ref{y_tensor}) respectively.}\label{figure5}
\end{figure}

\begin{figure}[h] \epsfig{width=3in,file= 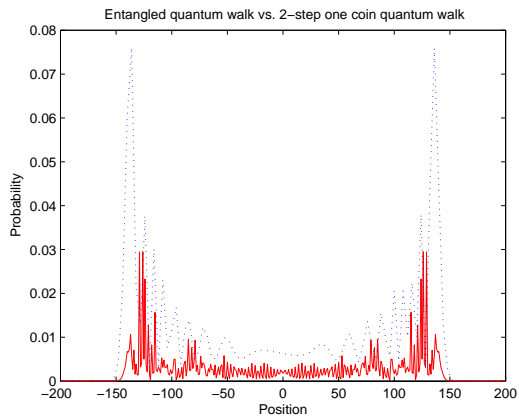}
\caption{For the red line plot, the coin initial state is given by
$|\Psi^+\rangle = \frac{1}{\sqrt{2}} (|00\rangle + |11\rangle)$.
Coin operator is given by Eq. (\ref{hadamard_tensor}) and shift
operator by Eq. (\ref{shift_operator_s2}).
For the blue dashed graph, coin initial state is given by 
$(\sqrt{0.85}|0\rangle_c -\sqrt{0.15} |1\rangle_c)$, coin operator is
the Hadamard operator (Eq. (\ref{hadamard_single})) as coin operator and
shift operator given by $|0\rangle \langle 0| \otimes \sum_i |i+2 \rangle \langle i|
+ |1\rangle \langle 1| \otimes \sum_i |i-2 \rangle \langle i|$.
In both cases, the number of steps is $100$.} \label{figure6}
\end{figure}

A consequence of the previous two properties of the quantum walk is
a sharper and narrower peak in the probability distribution around
position $|0\rangle$. Again, this may be of some advantage depending
on the application of the quantum walk (for example, less dispersion
around the most likely solution to the computational problem posed
in the two previous paragraphs).

The probability distributions for quantum walks in
Fig.~(\ref{figure3}) are very similar in structure to those of Fig.~
(\ref{figure2}), the only difference being the number of steps (200
as opposed to 100). For Fig.~(\ref{figure3}) the initial entangled
coin state is given by Eq.~(\ref{bell1}). Eq.~(\ref{hadamard_tensor})
is used as the coin operator in the red plot of Fig.~(\ref{figure3})
whereas Eq.~(\ref{y_tensor}) is the coin operator for the dotted blue
plot of Fig.~(\ref{figure3}). For $200$ steps the peaks on both
extreme zones are smaller than for $100$ steps, the reason being
the increased number of small probabilities that correspond to
those regions between the extreme peaks and the central peak.
A wider region is covered in the case of $200$ steps than for
$100$ steps.

An examination of Figs.~(\ref{figure4}) and (\ref{figure5}) is
straightforward, as their bulk properties closely resembling those
of Fig.~(\ref{figure2}) and Fig.~(\ref{figure3}).
The probability distributions in Fig.~({\ref{figure4}) and
were computed using Eq.~(\ref{bell2}) as the
initial coin state and the same initial conditions and shift
operators as for Fig.~(\ref{figure2}) and Fig.~(\ref{figure3}).
The number of steps is $100$ for red and dotted blue plots and
$200$ for starred yellow and black dotted plots.  
Coin operator for red and starred yellow plots is given by
Eq. (\ref{hadamard_tensor}), while coin operators for dotted
blue and dotted black plots is given by Eq. (\ref{y_tensor}) respectively.
The \lq three-peak zones' feature is again evident. Furthermore, the
bulk properties of the probability distributions are highly
invariant to changes in coin operators (there is a slight difference
in the probability distribution value at the origin). In both cases
the probability distribution value at the origin is much larger than
in the classical random walk case. A similar discussion having
Eq. (\ref{bell3}) as coin initial state, and using the same colors
as in Fig.~(\ref{figure4}) to refer to coin operators
and number of steps, applies to Fig.~(\ref{figure5}).

In order to further motivate the richness of quantum walks with
entangled coins, we present the graph shown in Fig.~(\ref{figure6}, red line plot)
computed using Eq. (\ref{bell1}) as the initial state of the coin, Eq.~
(\ref{hadamard_tensor}) as the coin operator and Eq.~ (\ref{shift_operator_s2})
as the shift operator. This graph closely resembles that of a 2-step
quantum walk Fig.~(\ref{figure6}, dotted blue plot) with initial state 
$(\sqrt{0.85}|0\rangle_c -\sqrt{0.15} |1\rangle_c) \otimes |0\rangle_p$ \cite{tregenna03},
Hadamard operator (Eq. (\ref{hadamard_single})) as coin operator and
shift operator given by
$|0\rangle \langle 0| \otimes \sum_i |i+2 \rangle \langle i| +
|1\rangle \langle 1| \otimes \sum_i |i-2 \rangle \langle i|$
(the number of steps in both walks is 100). However, the graph
corresponding to the quantum walks with a maximally entangled
coin has no parity restriction, as opposed to the 2-step quantum walk,
and this explains the higher probability values for the 2-step quantum walks.

As opposed to the previous cases (Figs. \ref{figure2} -
\ref{figure5}) in which the walker remains static when the quantum
coin state component is either $|01\rangle$ or $|10\rangle$, in this
case the walker is forced to jump either one or two steps, depending
on the components of the coin state. As can be seen in
Fig.~(\ref{figure6}), the behaviour of the quantum walk dramatically
changes as a consequence of the change in the shift operator. In
this case, constructive interference takes place not only in certain
areas of the graph (as is the case with the \lq three peak zones'
property) but in a wider region. Indeed, this walk bears a
resemblance to a quantum walk using a single walker and a single
Hadamard coin \cite{kempe04}.

Finally we would like to emphasize that in stark contrast to the
probability distributions of the classical case in which only
certain walker positions have a probability different from zero,
namely those positions whose parity is that of the total number of
steps, in the quantum cases presented in this paper we observe no
such constraint on the numerical data produced.
As stated in the introduction, the dynamics of
classical random walks can remove the parity constraint
by permitting the use of \lq rest sites' at the expense of
varying the amount of correlation between the coins.


\section{Quantum Walks using coins with different values of entanglement}

In order to compare the properties of quantum walks with coins
having different degrees of entanglement, we present in this
section several probability distributions computed using bipartite
coins. The graphs of those probability distributions are shown in
Figs. (\ref{figure7} - \ref{figure9}). All graphs shown in
Figs. (\ref{figure7} - \ref{figure9}) were computed
using Eq. (\ref{hadamard_tensor}) as coin operator and
Eq. (\ref{shift_operator_two_spins}) as shift operator.
The initial position state in all cases is the origin,
i.e. $|0\rangle_p$.

The probability distribution presented in Fig. (\ref{figure7})
shows a typical skewed (asymmetric) behaviour in quantum walks.
The graph is produced using the bipartite quantum state
$|\theta_0\rangle = {1 \over 2}(|0\rangle + |1\rangle)(|0\rangle + |1\rangle)$
as coin initial state (from Eq. (\ref{von_neumann}),
$E(|\theta_0\rangle) = 0$ so $|\theta_0\rangle$ is non-entangled).
This graph resembles the behaviour of
a quantum walk presented in \cite{brunetal203} using a coin
in initial state $|00\rangle$ ($|RR\rangle$ in their notation).

Let us now focus on the behaviour of the quantum walk
shown in Fig. (\ref{figure8}), which was produced
using a partially entangled initial coin state.
The coin was initialized in the state
$|\theta_1\rangle= {1 \over 2}|00\rangle + {1 \over 2}|01\rangle
+ {{\sqrt{3}-1} \over 4}|10\rangle +
{{\sqrt{3}+1} \over 4}|11\rangle$. Again, using Eq. (\ref{von_neumann}),
we find that $E(|\theta_1\rangle) = 0.5$, i.e. $|\theta_1\rangle$ is partially
entangled. 

We can see that an immediate effect of an entangled
coin initial state is the development of a third
peak, in the case a peak on the LHS of the graph.
This third peak reduces the skewness of the probability
distribution computed with a non-entangled coin
initial state (Fig. (\ref{figure9})).

\begin{figure}[h] \epsfig{width=3in,file= 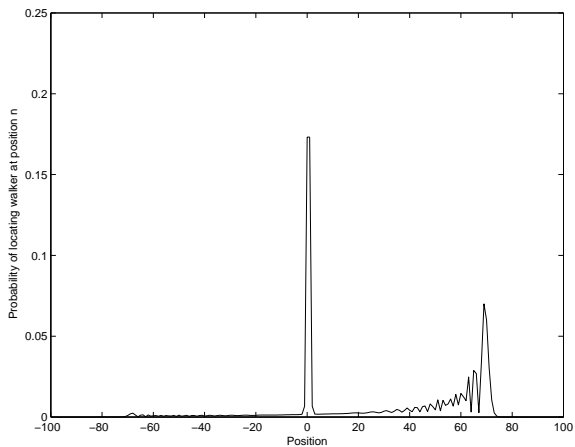}
\caption{Quantum walk computed with a coin initialized
in the state ${1 \over 2}(|0\rangle + |1\rangle)(|0\rangle + |1\rangle)$,
i.e. a non-entangled state with real coefficients. 100 steps,
coin and shift operators given by Eqs. (\ref{hadamard_tensor}) and
(\ref{shift_operator_two_spins}) respectively.} \label{figure7}
\end{figure}

\begin{figure}[h] \epsfig{width=3in,file= 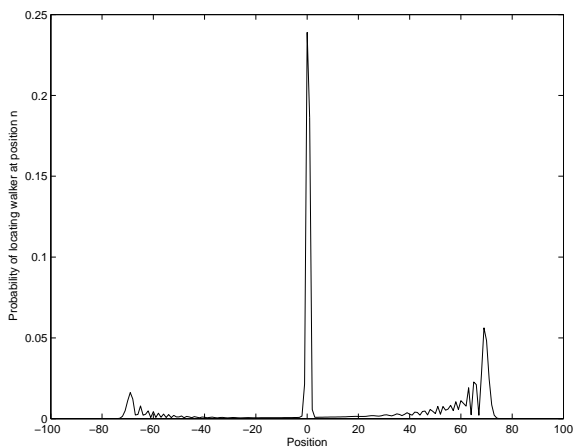}
\caption{Quantum walk computed with a coin initialized
in the state ${1 \over 2}|00\rangle + {1 \over 2}|01\rangle
+ {{\sqrt{3}-1} \over 2}|10\rangle + {{\sqrt{3}+1} \over 2}|11\rangle$,
i.e. a partially-entangled state with real coefficients. 100 steps,
coin and shift operators given by Eqs. (\ref{hadamard_tensor}) and
(\ref{shift_operator_two_spins}) respectively. Note that the
entanglement of the coin initial state reduces the assymetry of 
the graph by creating a new third peak.} \label{figure8}
\end{figure}

\begin{figure}[h] \epsfig{width=3in,file= 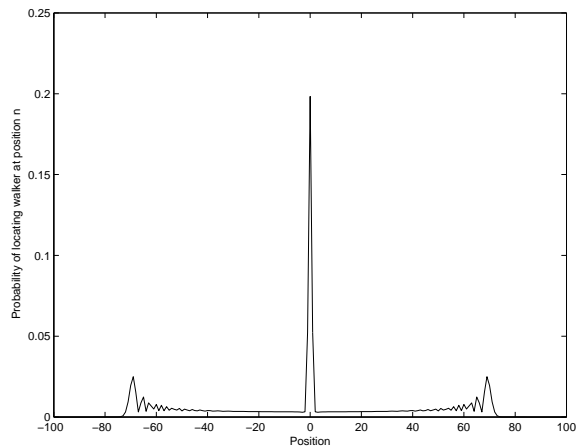}
\caption{Coin initial state is $\frac{1}{2}(|0\rangle + i|1\rangle) 
(|0\rangle + i|1\rangle)$. 100 steps, coin and shift operators given
by Eqs. (\ref{hadamard_tensor}) and (\ref{shift_operator_two_spins})
respectively. The use of complex coefficients in the coin initial
state delivers a symmetric probability distribution very similar
to those shown in Figs.(\ref{figure2} - \ref{figure5}).
} \label{figure9}
\end{figure}

\begin{figure}[h] \epsfig{width=3.5in,file= 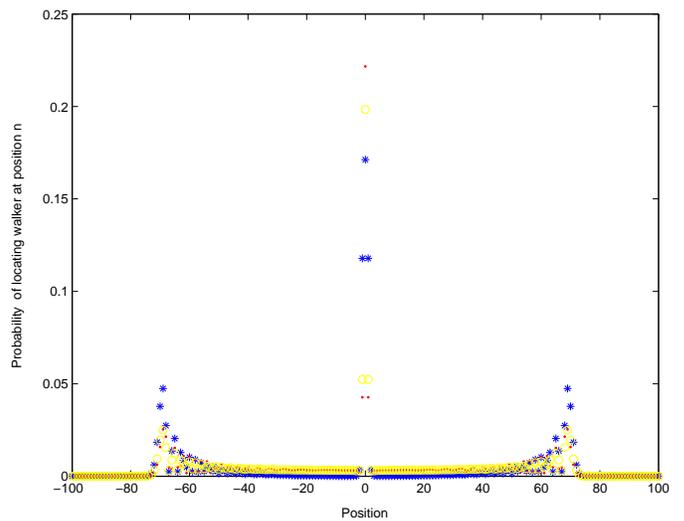}
\caption{ The probability distribution computed with coin
${1 \over \sqrt{2}}(|00\rangle + |11\rangle)$
corresponds to the blue starred graph. Probability distributions
computed with coins ${1 \over \sqrt{2}}(|00\rangle - |11\rangle)$
and ${1 \over 2}|00\rangle + {i \over 2}|01\rangle +
{i \over 2}|10\rangle - {1 \over 2}|11\rangle$
are shown in red dots and yellow circles respectively. All graphs
were computed after 100 steps using Eq. (\ref{hadamard_tensor})
as coin operator and Eq. (\ref{shift_operator_two_spins}) as
shift operator.}
\label{figure10}
\end{figure}

Let us now compare Figs. (\ref{figure9}) and (\ref{figure10})
with the red plot from Fig. (\ref{figure2}), created  with the maximally
entangled state  ${1 \over \sqrt{2}}(|00\rangle + |11\rangle)$
as initial coin state. We can see that the increasing
use of entanglement provides a greater degree of symmetry
to the resulting probability distribution.

Now, if we expand the properties of initial conditions by
allowing coins to be initialized in states with complex
coefficients, we would then obtain probability distributions
that would be similar to those of quantum walks with maximally
entangled coins with real coefficients.

For example, we show in Fig. (\ref{figure9}) the position
probability distribution computed with initial coin state
$\frac{1}{2}(|0\rangle + i|1\rangle) (|0\rangle + i|1\rangle)$.
Fig. (\ref{figure9}) bears a striking resemblance to those
of Figs. (\ref{figure2} - \ref{figure5}). However, even though
{\it qualitatively} a three peaked structure is evident, {\it quantitatively}
it differs considerably. To illustrate this numerical difference,
we appeal to Fig. (\ref{figure10}), which compares three different
quantum walks. The probability distribution computed with coin
${1 \over \sqrt{2}}(|00\rangle + |11\rangle)$
corresponds to the blue starred points on the graph, while
the probability distributions computed with
coins  ${1 \over \sqrt{2}}(|00\rangle - |11\rangle)$
and ${1 \over 2}|00\rangle + {i \over 2}|01\rangle +
{i \over 2}|10\rangle - {1 \over 2}|11\rangle$ are depicted
using red dots and yellow circles respectively. Thus the entanglement 
of the initial coin state helps to both tune up and tune down
the ratio of the central peak to the side peaks.

Numerical values show that entanglement plays an
active role in the actual probability of finding
the walker in a certain position. For example, consider
walker positions 60-70. The highest values in this region
are attained by the probability distribution computed
with coin ${1 \over \sqrt{2}}(|00\rangle + |11\rangle)$.
In a different region, that of the central peak,
the probability distribution of the non-entangled coin 
initial state ${1 \over 2}|00\rangle + {i \over 2}|01\rangle +
{i \over 2}|10\rangle - {1 \over 2}|11\rangle$
is between those values produced by the
two probability distributions obtained
by computing quantum walks with maximally
entangled states.

Another example of a symmetric graph produced using
coins in non-entangled states with complex coefficients
has been presented by Inui and Konno in \cite{konno04}.
This symmetric graph was produced using a coin initialized
in the state ${i \over 2}|00\rangle + {i \over 2}|01\rangle +
{1 \over 2}|10\rangle + {1 \over 2}|11\rangle$.

\section{Quantum walks with more than two maximally entangled coins}

An interesting property of using several entangled qubits
as coins is the fact that the number of coin and shift operators available for use
also increases. Consequently, several different position
probability distributions can be computed.

For example, in Fig.(\ref{figure11}, red plot) the graph of a 100-steps
quantum walk with the GHZ state ${1 \over \sqrt{2}}(|000\rangle + |111\rangle)$
as coin initial state is shown, the coin operator being given by 
${\hat H}^{\otimes 3}$ where $\hat{H}$ is Hadamard operator (Eq. (\ref{hadamard_single})),
and shift operator given by

\begin{equation}
\begin{split}
{\hat S}_{3a} = |000\rangle_{cc} \langle 000| \otimes \sum_i |i+1 \rangle_{pp}\langle i| \\
+ |001\rangle_{cc} \langle 001| \otimes \sum_i |i\rangle_{pp} \langle i| \\
+ |010\rangle_{cc} \langle 010| \otimes \sum_i |i\rangle_{pp} \langle i|\\
+ |011\rangle_{cc} \langle 011| \otimes \sum_i |i\rangle_{pp} \langle i| \\
+ |100\rangle_{cc} \langle 100| \otimes \sum_i |i\rangle_{pp} \langle i|\\
+ |101\rangle_{cc} \langle 101| \otimes \sum_i |i\rangle_{pp} \langle i| \\
+ |110\rangle_{cc} \langle 110| \otimes \sum_i |i\rangle_{pp} \langle i|\\
+ |111\rangle_{cc} \langle 111| \otimes \sum_i |i-1\rangle_{pp} \langle i|
\end{split}
\label{three_qubit_shift_operator_a}
\end{equation}
which has a $4$-peak probability distribution. However, changing the shift operator to

\begin{equation}
\begin{split}
{\hat S}_{3b} = |000\rangle_{cc} \langle 000| \otimes \sum_i |i+3 \rangle_{pp}\langle i| \\
+ |001\rangle_{cc} \langle 001| \otimes \sum_i |i+2\rangle_{pp} \langle i| \\
+ |010\rangle_{cc} \langle 010| \otimes \sum_i |i+1\rangle_{pp} \langle i|\\
+ |011\rangle_{cc} \langle 011| \otimes \sum_i |i\rangle_{pp} \langle i| \\
+ |100\rangle_{cc} \langle 100| \otimes \sum_i |i\rangle_{pp} \langle i|\\
+ |101\rangle_{cc} \langle 101| \otimes \sum_i |i-1\rangle_{pp} \langle i| \\
+ |110\rangle_{cc} \langle 110| \otimes \sum_i |i-2\rangle_{pp} \langle i|\\
+ |111\rangle_{cc} \langle 111| \otimes \sum_i |i-3\rangle_{pp} \langle i|
\label{three_qubit_shift_operator_b}
\end{split}
\end{equation}
results in the blue plot of Fig. (\ref{figure11}) which has no such readily evident
peak structure.

\begin{figure}[h] \epsfig{width=3.5in,file= 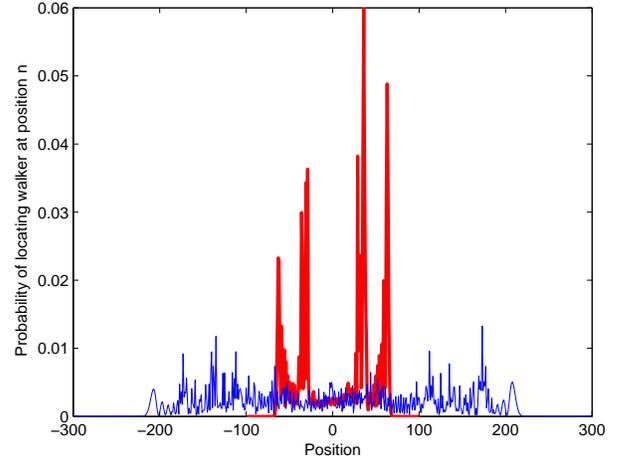}
\caption{Position probability distributions for two quantum walks on a line
with GHZ state ${1 \over \sqrt{2}}(|000\rangle + |111\rangle)$  as initial tripartite coin state
and coin operator ${\hat H}^{\otimes 3}$. The red plot was computed  using the shift operator in
Eq. (\ref{three_qubit_shift_operator_a}) and the blue plot using the shift operator in
Eq. (\ref{three_qubit_shift_operator_b}). While the red plot shows an evident
4-peak structure, the blue plot does not present such a behaviour.}
\label{figure11}
\end{figure}

The potential richness of quantum walks increases when taking into consideration
graphs of more than one dimension (efforts to understand
the properties of quantum walks on graphs are presented in \cite{aharonov01},
 while a proposal for a physical realization of a 2-dimensional
quantum walk is given in \cite{roldan05}). For example, Fig.
(\ref{figure12}) shows the peak structure of a 50-step quantum walk on a graph
with initial state given again by ${1 \over \sqrt{2}}(|000\rangle + |111\rangle)$,
coin operator given by $\hat{H}^{\otimes 3}$ 
and shift operator given by Eq. (\ref{three_qubit_shift_operator_c})

\begin{equation}
\begin{split}
{\hat S}_{EC} = |000\rangle_{cc} \langle 000| \otimes \sum_i |i+1,j \rangle_{pp}\langle i,j| \\
+ |001\rangle_{cc} \langle 001| \otimes \sum_i |i,j\rangle_{pp} \langle i,j| \\
+ |010\rangle_{cc} \langle 010| \otimes \sum_i |i,j+1\rangle_{pp} \langle i,j|\\
+ |011\rangle_{cc} \langle 011| \otimes \sum_i |i,j\rangle_{pp} \langle i,j| \\
+ |100\rangle_{cc} \langle 100| \otimes \sum_i |i,j\rangle_{pp} \langle i,j|\\
+ |101\rangle_{cc} \langle 101| \otimes \sum_i |i,j-1\rangle_{pp} \langle i,j| \\
+ |110\rangle_{cc} \langle 110| \otimes \sum_i |i,j\rangle_{pp} \langle i,j|\\
+ |111\rangle_{cc} \langle 111| \otimes \sum_i |i-1,j\rangle_{pp} \langle i,j|
\label{three_qubit_shift_operator_c}
\end{split}
\end{equation}


\begin{figure}[h] \epsfig{width=3.5in,file= 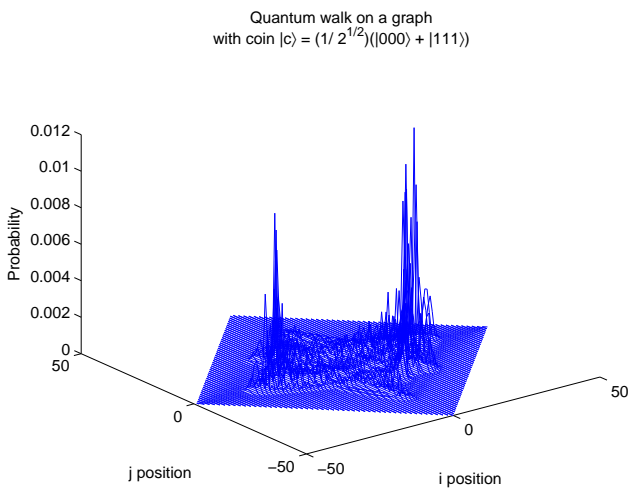}
\caption{Position probability distribution of a quantum walk
on a $2$-dimensional graph computed with coin
initial state ${1 \over \sqrt{2}}(|000\rangle + |111\rangle)$
and shift operator given by Eq. (\ref{three_qubit_shift_operator_c}).
The number of steps is 50. The graph has $2$ high peak regions and several other
small peaks in the central region.}
\label{figure12}
\end{figure}

\section{Conclusions}

We have studied quantum walks with maximally entangled coin initial
states and have compared their behaviour with that of a classical
random walk with a maximally correlated pair of coins as well as
that of quantum walks with different degrees of entanglement. The
probability distributions of such quantum walks have particular
forms which are markedly different from the probability distributions
of maximally correlated classical random walks. As for the single coin
and entangled coins quantum walks, by changing the
shift operator in the entangled case, one can generate a multitude of 
different probability distributions, some of which clearly differ
from their single coin quantum walk counterparts.

The basic \lq three peak zone' form is reproduced for a number
of different entangled coin operators. In this case, the probability of
finding the walker in the most likely position also appears to be
higher when performing a quantum walk with a maximally entangled
coin than when computing its classical counterpart (classical random
walk with maximally correlated coin pair).

We have also considered how the \lq three peak zone' form can
also be produced by a quantum walk with coins using different
initial conditions, i.e. a non-entangled coin with complex
coefficients. Even though the shape of both probability
distributions is similar, the quantum walks with maximally
entangled coins have a different quantitative behaviour (higher
or lower peaks, depending on the specific maximally entangled
coin used). Entanglement allows symmetry in our probability
distributions without using complex coefficients in initial
coin states.


\section{Acknowledgments}

S.E. Venegas-Andraca gratefully acknowledges useful discussions with
N. Paunkovi\'{c}. This work was supported by CONACyT-M\'{e}xico
(scholarship 148528) and the UK Engineering and Physical Sciences
Research Council.


\end{document}